\documentclass[10pt,journal,compsoc]{IEEEtran}

\usepackage[binary-units=true,per-mode=repeated-symbol]{siunitx}

\ifCLASSOPTIONcompsoc
  \usepackage[nocompress]{cite}
\else
  \usepackage{cite}
\fi
\ifCLASSINFOpdf
  \usepackage[pdftex]{graphicx}
\else
  \usepackage[dvips]{graphicx}
\fi
\usepackage{amsmath}
\usepackage{algorithmic}

\usepackage{array}

\ifCLASSOPTIONcompsoc
 \usepackage[caption=false,font=footnotesize,labelfont=sf,textfont=sf]{subfig}
\else
 \usepackage[caption=false,font=footnotesize]{subfig}
\fi
\usepackage{url}

\usepackage{hyperref}
\hypersetup{
  colorlinks=true,
  urlcolor=blue,
  linkcolor=black,
  citecolor=blue,
}

\begin{document}
%

\title{Public Release and Validation of \\ SPEC CPU2017 PinPoints}

%
%
%
%

\author{
Haiyang~Han\IEEEauthorrefmark{2}
and~Nikos~Hardavellas\IEEEauthorrefmark{2}\IEEEauthorrefmark{3}
\IEEEcompsocitemizethanks{\IEEEcompsocthanksitem
\IEEEauthorrefmark{2}Department of Electrical and Computer Engineering and
\IEEEauthorrefmark{3}Department of Computer Science,
Northwestern University, Evanston, IL, 60208.\protect\\
E-mail: haiyang.han@u.northwestern.edu, nikos@northwestern.edu.}%
}

\IEEEtitleabstractindextext{%
\begin{abstract}
Phase-based statistical sampling methods such as SimPoints have proven to be effective at
dramatically reducing the long time for architectural simulators to run large workloads
such as SPEC CPU2017. However, generating and validating them is a long and tenuous process.
While checkpoints of program phases, or ``pinballs'', of SPEC CPU2017 have been
collected by other researchers and shared with the research community, they are outdated
and produce errors when used with the latest versions of the Sniper architectural simulator.
To facilitate our own research as well as contribute to the community, we collect and validate
our own pinballs for the SPEC CPU2017 SPECspeed suite and
release them to the public domain. In this work we document our methodology, the hardware
and software details of the collection process, and our validation results.
In terms of CPI, our pinballs have an average error rate of 12\% when compared with the
native whole-program benchmark execution.
\end{abstract}

}

\maketitle

\IEEEdisplaynontitleabstractindextext

%
\IEEEpeerreviewmaketitle

\IEEEraisesectionheading{\section{Introduction}\label{sec:introduction}}
\IEEEPARstart{}
{In} architecture research, evaluating novel ideas before they are physically implemented
requires modeling them on simulators that execute a wide range of representative benchmarks.
Released in 2017, the SPEC (Standard Performance Evaluation Corporation) CPU 2017 benchmark suite~\cite{cpu2017} has been a popular tool in computer architecture for such tasks.
However, the increased dynamic instruction counts and large memory footprints of CPU2017 over its predecessors have led to unrealistically long simulation times. The increasingly complex modeling of novel architectures, possibly with high core counts, sophisticated cache and Network-on-Chip protocols, and emerging materials further exacerbate the problem---some benchmarks take months to simulate to completion. To combat this problem, phase-based statistical sampling methods like SimPoints~\cite{sherwood2002automatically} were developed, in which simulating multiple short representative regions can predict the behavior of the whole application, significantly reducing the total simulation time. The Intel PinPoints~\cite{patil2004pinpointing} tool set automates the region-finding process by using the program instrumentation tools Pin~\cite{luk2005pin} and PinPlay~\cite{patil2010pinplay}. Each generated representative region, also called a pinball, can be replayed on simulators such as Sniper~\cite{carlson2011etloafsaapms} and ZSim~\cite{sanchez2013zsim} and shared among researchers, liberating them from the need to collect sampled regions.

\textit{Pinballs} of CPU2017 are already available online~\cite{wu2018hot, cpu2017pinballs} from other researchers. However, our own attempts at using these pinballs with the latest version of the Sniper simulator were unsuccessful as various errors would end simulations before performance statistics could be collected. Thus we decided to collect our own pinballs of the CPU2017 SPECspeed benchmarks, not only to use in our own research, but also to provide an alternate set of CPU2017 pinballs that researchers from around the world might find useful. The link to our pinballs repository can be found on our lab's website, PARAG@N, under ``Pinballs'' in the Artifacts section at \href{http://paragon.cs.northwestern.edu/\string#Artifacts}{http://paragon.cs.northwestern.edu/\#Artifacts}. This work provides the details of our pinball-collecting process, pinball statistics, and validation of representability against whole programs in terms of cycles per instruction (CPI).

Our generated pinballs achieve an average absolute error rate of 12\% when comparing their predicted CPIs with those of dynamically linked native applications in the SPEC CPU2017 SPECspeed suite. We also find that for applications with high CPI prediction error rates, comparing against statically linked applications can reduce their error rate by an average of 29.7\%, bringing the average absolute error rate across the entire SPEC CPU2017 SPECspeed suite down to 8\%.

\section{Background}\label{sec:background}

\subsection{SPEC CPU2017}\label{sec:background/cpu2017}

SPEC CPU2017 is a collection of 43 benchmarks categorized into four suites: SPECspeed Integer, SPECspeed Floating Point, SPECrate Integer, and SPECrate Floating Point. The SPECspeed benchmarks are mainly used for measuring execution time by always running one copy of a benchmark while SPECrate measures throughput by running multiple concurrent copies of each benchmark. CPU2017 also provides three input sizes: \textit{test}, \textit{train}, and \textit{ref}. Only the \textit{ref} input size can be used when reporting time metrics.

\subsection{Pinballs}\label{sec:background/pinballs}
Pin~\cite{luk2005pin} from Intel is a dynamic binary instrumentation framework. Tools created using Pin can be used to analyze user space applications at the instruction set architecture (ISA) level. It does not require recompiling the source code of a target application because the instrumentation is performed at run time. The Program Record/Replay Toolkit~\cite{patil2010pinplay}, or PinPlay is a set of tools built using Pin that support the logging and replaying of an entire program or a part of it. Running PinPlay's logger on a program produces a pinball, which can either be replayed using PinPlay itself or fed into a simulator. A pinball is a collection of files that contain information about a program like initial memory and register states, register states before and after system calls, etc. When replayed, the pinball guarantees repeatable and deterministic behavior. A large pinball of a whole program can be sliced into regional pinballs.

SimPoints~\cite{perelman2003using} is a tool that uses statistical sampling to capture multiple representative simulation regions of a large program. The simulation points produced by SimPoints contain the bulk of information about a program's execution and can be used to accurately model the run time behavior of the whole program. Each simulation point is associated with a weight that roughly represents the relative frequency by which the corresponding phase executes in the program. Because only small phases of the program are captured, simulation time with SimPoints is greatly reduced.

\begin{figure}[!t]
    \centering
    \includegraphics[width=\columnwidth]{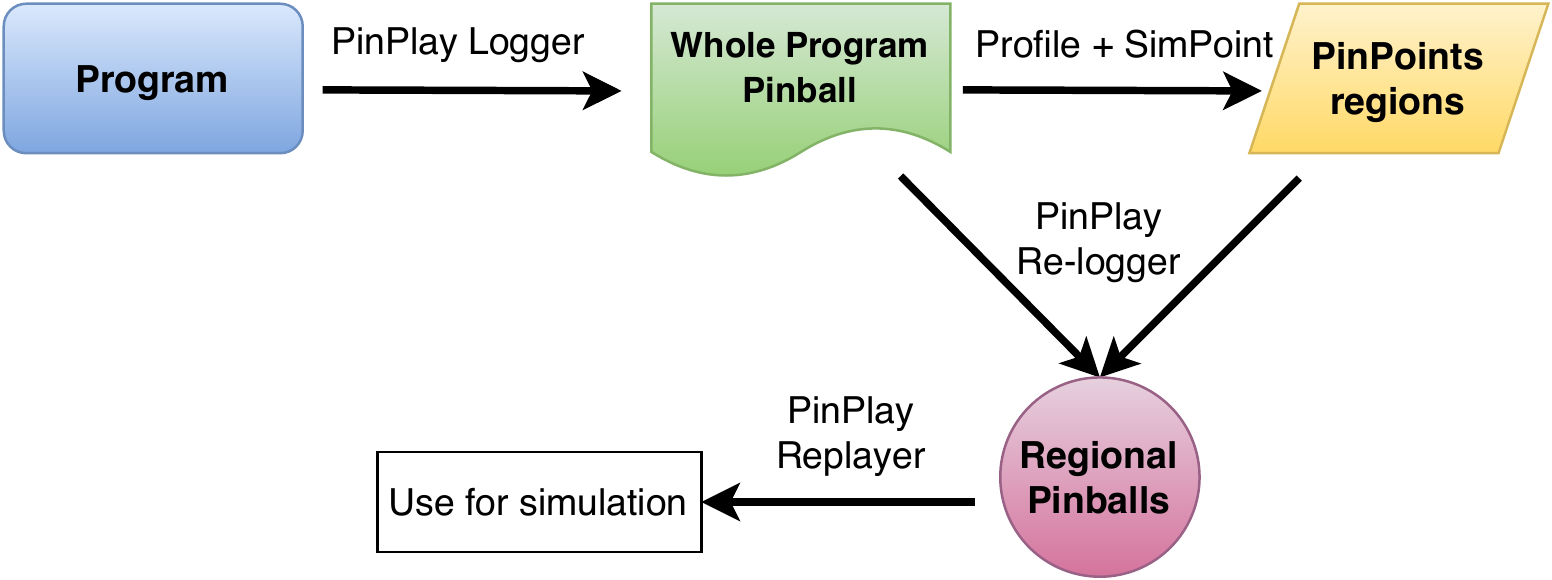}
    \caption{PinPoints workflow.}
    \label{fig:pinpoints_flow}
\end{figure}

PinPoints~\cite{patil2004pinpointing} combines Pin with SimPoints. It uses the whole program pinball recorded with PinPlay as input to SimPoints and produces simulation points in the form of regional pinballs. Figure~\ref{fig:pinpoints_flow} shows the workflow for capturing regional pinballs. There are a number of advantages with using pinballs produced by PinPoints. They are OS independent, provide reproducible and deterministic simulation results, and can be shared among researchers.

\subsection{ELFies}\label{sec:background/elfies}
ELFies~\cite{patil2021elfies} or \textit{pinballs2elf} is a tool-chain that converts a pinball into an ELF executable, which can be run natively on Linux and without extra overhead. This is useful when validating the regions produced by SimPoints. To validate SimPoints, one needs to compute the error rate by calculating (1) the weighted average of CPI of the regional pinballs, and (2) the CPI of the whole program pinball. Typically, the CPIs are collected by running the pinballs through a simulator. While step 1 can be quite fast as the slices are relatively short, step 2 requires the simulation of the entire program, which can be unrealistically time consuming. In our experience, the fastest whole-program simulation in the CPU2017 suite using the Sniper architectural simulator finished in 28 days when running on a modern Intel-based server.
The ELF executables generated by ELFies significantly reduce the validation time by allowing one to execute regional pinballs natively and use hardware counters to calculate the CPI. In this way we are able to quickly validate our pinballs by comparing the CPI obtained by hardware performance counters of the original application and the ELFies generated from the regional pinballs.
\section{Methodology}\label{sec:methodology}

\begin{figure}[!t]
    \centering
    \includegraphics[width=0.7\columnwidth]{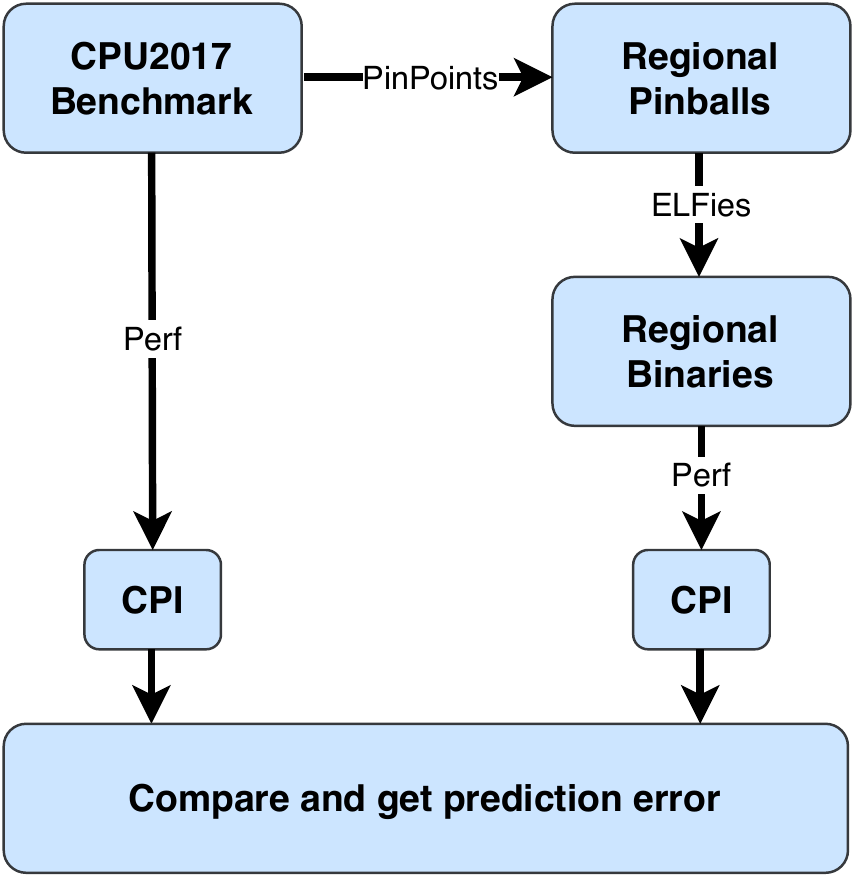}
    \caption{Validation workflow.}
    \label{fig:methodology_flow}
\end{figure}

\begin{figure*}[!ht]
    \centering
    \subfloat[Prediction errors (\%).]{\includegraphics[width=\columnwidth]{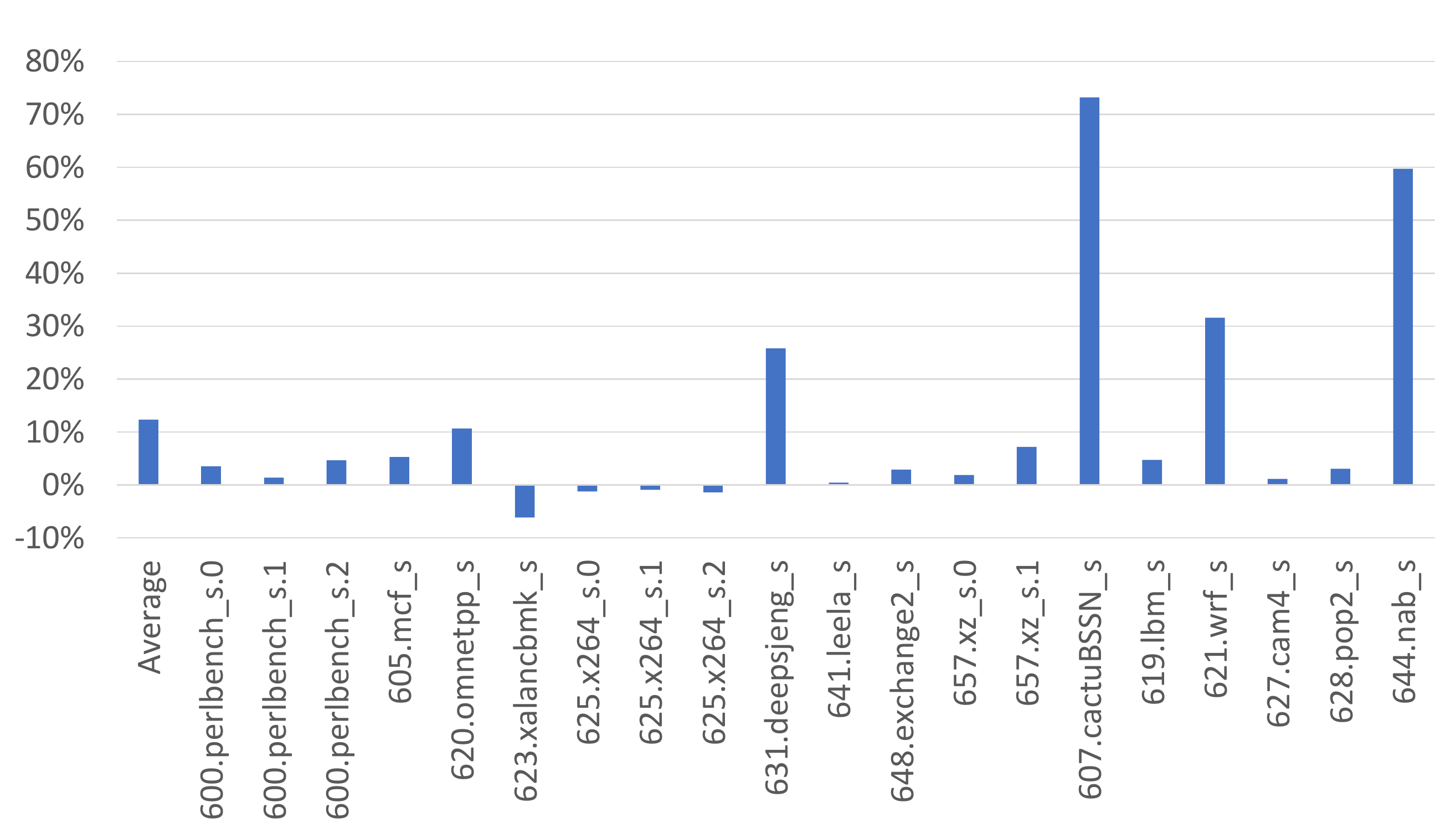}}
    \subfloat[CPI comparison.]{\includegraphics[width=\columnwidth]{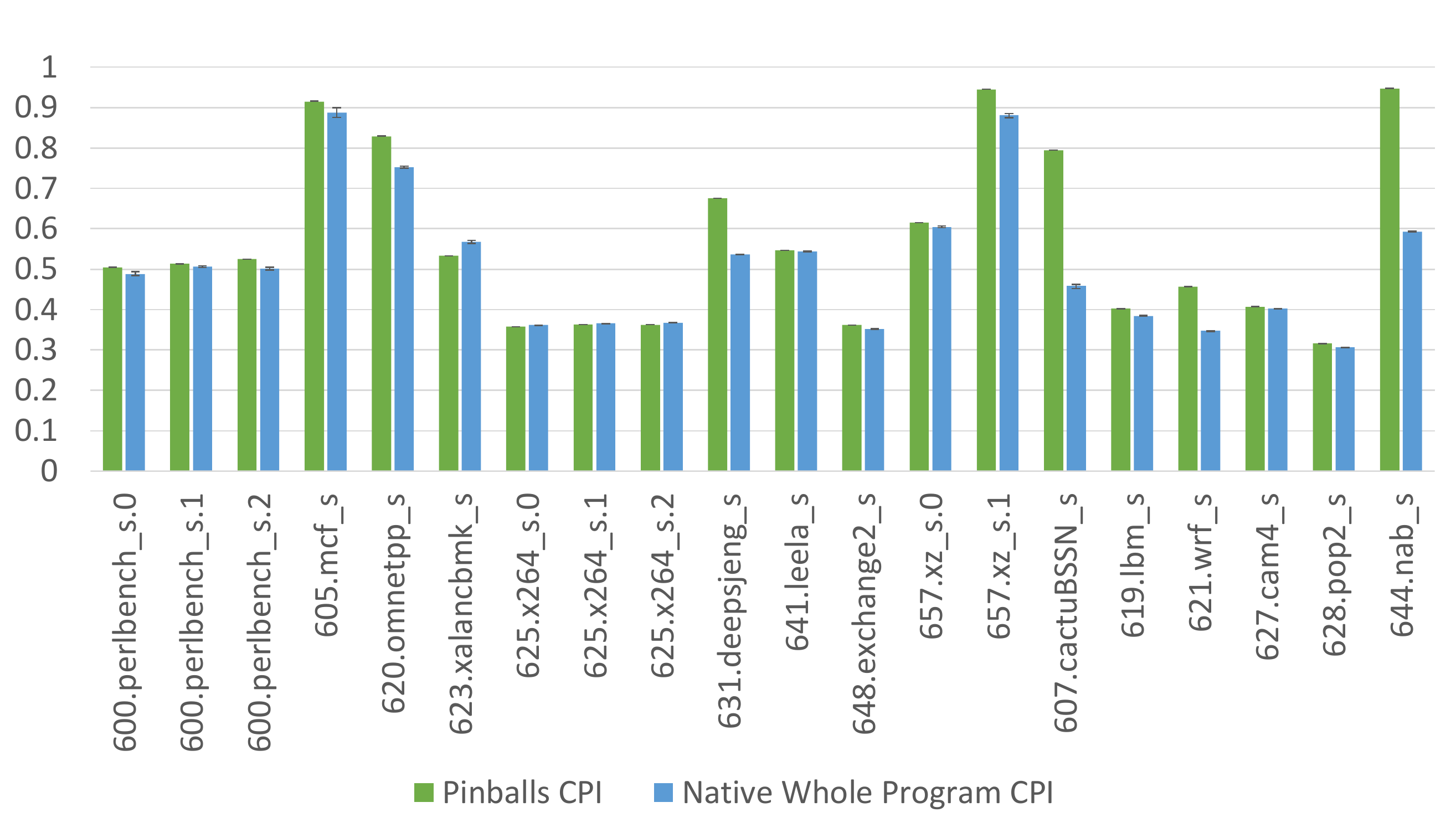}}
    \caption{Pinball validation results.}
    \label{fig:cpi_comparison}
\end{figure*}

\begin{table}[!t]
    \caption{CPU2017 SimPoints.}
    \label{fig:simpoints}
    \centering
    \begin{tabular}{lccr}
    \hline
    Benchmark          & \multicolumn{1}{p{0.2\columnwidth}}{\# of SimPoints} & \multicolumn{1}{p{0.2\columnwidth}}{90 Percentile SimPoints} & \multicolumn{1}{p{0.2\columnwidth}}{Instructions (billion)} \\ \hline
    \multicolumn{4}{l}{\textbf{SPEC Int}}                                                                       \\
    600.perlbench\_s.0 & 16              & 9                       & 1961.85                                    \\
    600.perlbench\_s.1 & 12              & 4                       & 1150.03                                    \\
    600.perlbench\_s.2 & 18              & 9                       & 1109.25                                    \\
    602.gcc\_s.0       & 19              & 6                       & 4721.74                                    \\
    602.gcc\_s.1       & 22              & 12                      & 1412.38                                    \\
    602.gcc\_s.2       & 23              & 11                      & 1350.69                                    \\
    605.mcf\_s         & 29              & 17                      & 4066.58                                    \\
    620.omnetpp\_s     & 3               & 2                       & 5951.38                                    \\
    623.xalancbmk\_s   & 23              & 18                      & 9226.90                                    \\
    625.x264\_s.0      & 25              & 16                      & 1522.16                                    \\
    625.x264\_s.1      & 20              & 14                      & 5515.79                                    \\
    625.x264\_s.2      & 15              & 9                       & 5560.37                                    \\
    631.deepsjeng\_s   & 6               & 5                       & 4848.04                                    \\
    641.leela\_s       & 21              & 13                      & 13728.15                                   \\
    648.exchange2\_s   & 19              & 15                      & 10596.43                                   \\
    657.xz\_s.0        & 18              & 10                      & 12910.01                                   \\
    657.xz\_s.1        & 17              & 11                      & 8018.37                                    \\
    Average            & 18              & 10.65                   & 5508.83                                    \\
    \hline
    \multicolumn{4}{l}{\textbf{SPEC FP}}                                                                        \\
    603.bwaves\_s.0    & 27              & 5                       & 59025.56                                   \\
    603.bwaves\_s.1    & 31              & 6                       & 55580.46                                   \\
    607.cactuBSSN\_s   & 31              & 6                       & 32636.33                                   \\
    619.lbm\_s         & 16              & 8                       & 18501.50                                   \\
    621.wrf\_s         & 27              & 20                      & 114200.20                                  \\
    627.cam4\_s        & 21              & 13                      & 38725.31                                   \\
    628.pop2\_s        & 20              & 12                      & 95140.64                                   \\
    644.nab\_s         & 16              & 7                       & 29067.15                                   \\
    649.fotonik3d\_s   & 23              & 10                      & 123075.22                                  \\
    654.roms\_s        & 30              & 23                      & 123075.22                                  \\
    Average            & 24.2            & 11                      & 61211.27                                   \\ \hline
    \end{tabular}
\end{table}

Figure~\ref{fig:methodology_flow} shows the entire workflow for our experiments. We use PinPoints (with Pin version 3.7) to generate regional pinballs of the SPEC CPU2017 SPECspeed benchmarks (both INT and FP) with the \textit{ref} input size. All benchmarks are compiled on an Intel\textsuperscript{\textregistered} Xeon\textsuperscript{\texttrademark} E5-2695 V3 processor (Haswell, 14 cores, \SI{2.3}{\giga\hertz}, \SI{32}{\kilo\byte} L1 cache, \SI{256}{\kilo\byte} L2 cache, \SI{35}{\mega\byte} last-level cache) running Linux kernel 4.18 using GCC 8. We disable the OpenMP compilation flags but keep all other optimization flags consistent with the recommended example provided by the CPU2017 distribution and target the 64-bit ISA. As for tuning the parameters of SimPoint, we use a \textit{maxk} value of 32 (32 maximum regions), a \textit{slice\_size} of 100 million instructions, and a \textit{warmup\_length} of 300 million instructions. We leave out some FP benchmarks because they could take months, as the logging and replaying process can incur as much as 200$\times$ slowdown compared to native execution~\cite{patil2015tutorial}. Table~\ref{fig:simpoints} summarizes the SimPoints and global dynamic instruction counts of the benchmarks we run. Note that the executions of \textit{600.perlbench\_s}, \textit{602.gcc\_s}, \textit{625.x264\_s}, \textit{657.xz\_s}, and \textit{603.bwaves\_s} have multiple steps, with each step taking in a different input. We consider each step as a separate benchmark, denoted by the number index suffixes after the dot in the \textbf{Benchmark} column. The \textbf{90 Percentile SimPoints} column shows the least number of SimPoints needed to reach a cumulative weight of 0.9 or more.

To evaluate the pinballs, we use ELFies to convert the region pinballs into native executables and use Linux \texttt{perf} to read hardware performance counters and calculate the CPIs of the individual regions. We then take measurements on a system with an Intel\textsuperscript{\textregistered} Core\textsuperscript{\texttrademark} i7-8700 processor (Skylake, 6 cores, \SI{3.2}{\giga\hertz}, \SI{32}{\kilo\byte} L1 cache, \SI{256}{\kilo\byte} L2 cache, \SI{8}{\mega\byte} last-level cache) and \SI{16}{\giga\byte} of DRAM, running Ubuntu 18.04 with a 4.18 Linux kernel.
Using a system for measurement that is different than the system used
for collecting the pinballs reflects the actual way that the pinballs we release will be used in the wild.

We measure each region or whole program instance ten times and calculate the average CPI.
The ELFies are generated with the hardware performance counters \texttt{PERF\_COUNT\_HW\_CPU\_CYCLES} and \texttt{PERF\_COUNT\_HW\_CPU\_INSTRUCTIONS} found in \texttt{/usr/include/linux/perf\_event.h}. To measure whole program performance, we use \texttt{perf record} and sample the \textit{cpu\_clk\_unhalted.thread} and \textit{inst\_retired.any} hardware counters. To minimize OS noise and DVFS effects, we turn off Intel Hyper-Threading, Turbo Boost, and SpeedStep, and run all experiments at the \texttt{init 3} runlevel. We use Sniper 7.4 with two versions of Pin (3.7 and 3.11) to test the generated pinballs and verify that they can run without errors.
We calculate the region predicted CPIs and prediction errors according to the equation:
\begin{center}
    $ predicted\_CPI = \displaystyle\sum\limits_{i=1}^{num\_simpoints}CPI_i \times weight_i $
    $ pred\_error = \frac{\lvert whole\_program\_CPI - predicted\_CPI \rvert}{whole\_program\_CPI} $
\end{center}
\begin{figure}[h]
    \centering
    \subfloat[Prediction errors (\%).]{\includegraphics[width=0.4\columnwidth]{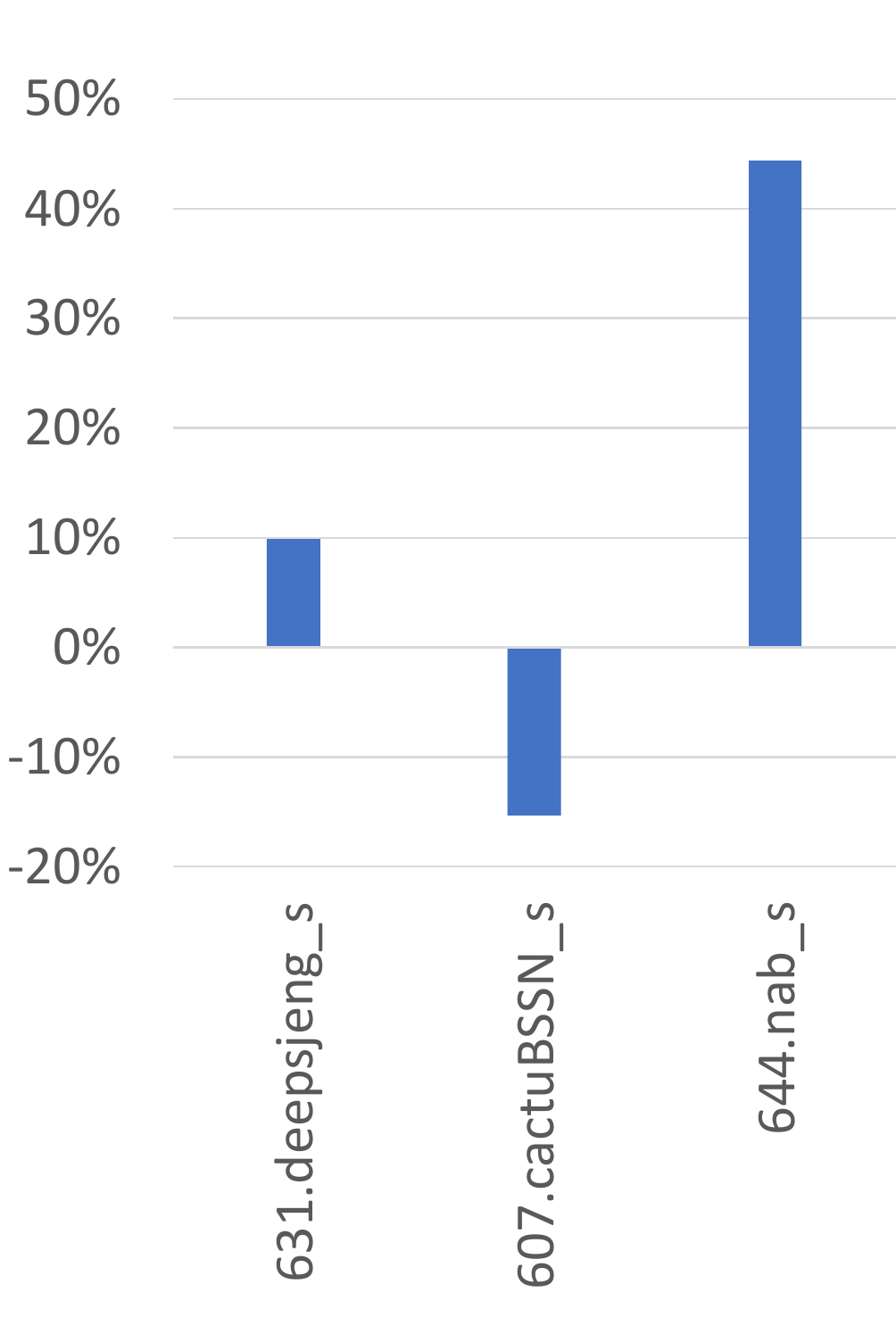}}
    \subfloat[CPI comparison.]{\includegraphics[width=0.5\columnwidth]{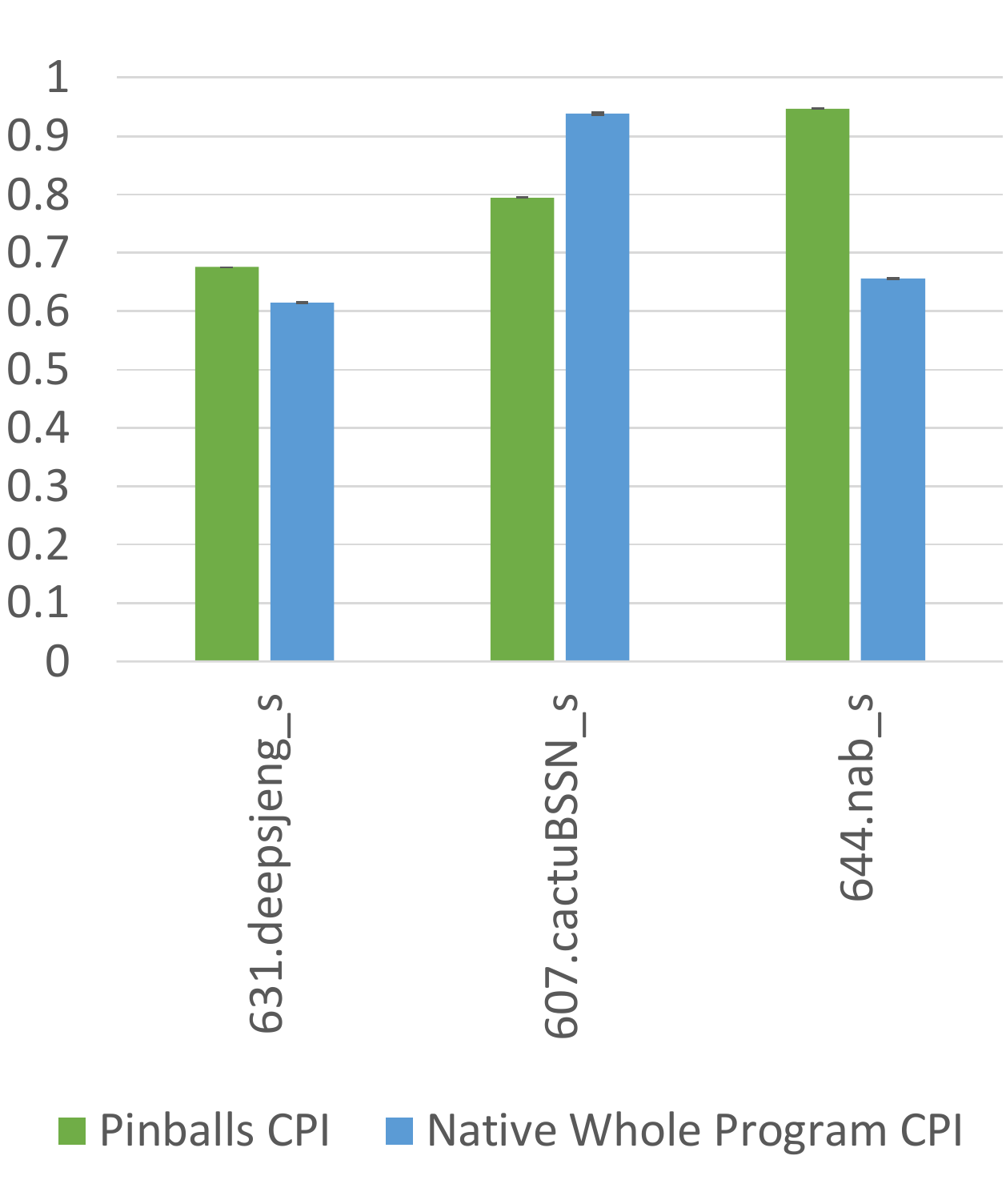}}
    \caption{Validation results for statically linked binaries.}
    \label{fig:cpi_comparison_static}
\end{figure}

\section{Results}\label{sec:results}

Figure~\ref{fig:cpi_comparison} shows the prediction errors and CPI comparisons of our pinballs against native runs of the benchmarks. The average absolute prediction error rate across all benchmarks is 12\%. For most benchmarks, our pinballs represent native execution well, and we can expect the pinballs to reasonably represent the performance characteristics of their native counterparts, while significantly reducing simulation time.

Benchmarks \textit{631.deepsjeng\_s}, \textit{607.cactuBSSN\_s}, \textit{621.wrf\_s}, and \textit{644.nab\_s} have error rates above 25\%. We posit that this is caused by the difference between dynamically linked libraries on the machine used to collect the pinballs and the machine used to collect runtime execution CPIs. To investigate this, we statically link \textit{631.deepsjeng\_s}, \textit{607.cactuBSSN\_s}, and \textit{644.nab\_s}, and compare the CPIs calculated by the pinball regions with the whole-program CPIs of the statically compiled benchmarks.
We show the prediction errors and CPI comparisons in Figure~\ref{fig:cpi_comparison_static}. Statically linking \textit{621.wrf\_s} requires \texttt{libgfortran.a}, which is not available in our version of RHEL, thus we exclude this application from further investigation. Using statically linked binaries reduces the absolute CPI error of these three benchmarks by an average of 29.7\%. This also brings the average error across the entire benchmark suite down to less than 8\%.

Unfortunately, licensing restrictions prevent us from releasing statically linked versions of \textit{631.deepsjeng\_s}, \textit{607.cactuBSSN\_s}, and \textit{644.nab\_s}.
Users of our pinballs (or any other pinball releases, for that matter) should be aware of the relatively high errors when using dynamically linked versions of these three benchmarks. Alternatively, users can statically compile their own versions of these three benchmarks on their own platforms and recollect the corresponding pinballs, or at least verify the pinballs' accuracy on their machines by comparing the calculated CPI with our published data.

Although further in-depth tuning of PinPoint parameters to generate more accurate pinballs for these four benchmarks is possible, that is a topic for future work.
\section{Conclusions}\label{sec:conclusion}

In this work we announce to the computer architecture community the public release of validated pinballs for SPEC CPU2017 SPECspeed benchmarks, and share the details of our pinball-collecting process, statistics, and validation results.
Our validation shows that the average absolute CPI error rate of our pinballs is 12\% for dynamically linked benchmarks.
We also discover that differences exist between dynamically linked and statically linked benchmarks when their CPIs are compared against the CPI results predicted from pinballs. For benchmarks that exhibit high errors when compiled dynamically, we find that compiling them statically can reduce the CPI error rate by 29.7\%. In particular, when statically linking \textit{631.deepsjeng\_s}, \textit{607.cactuBSSN\_s}, and \textit{644.nab\_s}, and dynamically linking all other benchmarks, we estimate an average absolute CPI error of less than 8\% across the entire SPEC SPU2017 suite.
The link to our pinballs repository can be found on our lab's website, PARAG@N, under ``Pinballs'' in the Artifacts section at \href{http://paragon.cs.northwestern.edu/\string#Artifacts}{http://paragon.cs.northwestern.edu/\#Artifacts}.

\section*{Acknowledgements}
This work was partially funded by NSF award CCF-1453853.


%





\ifCLASSOPTIONcaptionsoff
  \newpage
\fi



\bibliographystyle{IEEEtran}
\bibliography{IEEEabrv,references.bib}
\end{document}